%% file: main.tex
\documentclass[10pt]{IEEEtran}

\usepackage{tikz}
\usepackage{amsmath}
\usepackage{xspace}
\usepackage{enumitem}
\usepackage{amsfonts}
\usepackage{graphicx}
\usepackage{tikz}
\usepackage{amsmath}
\usepackage{times}
\usepackage{subfigure}
\usepackage{caption} 
\usepackage{algorithm}
\usepackage{algorithmic}
\usepackage{tabularx}
\usepackage{xspace}
\usepackage{booktabs}
\usepackage{multirow}
\usepackage{enumitem}
\usepackage{makecell}
\setlist{nosep}
\usepackage{url}
\usepackage{orcidlink}

\usepackage{amsmath}
\usepackage{amssymb}
\usepackage{mathtools}
\usepackage{amsthm}

\hypersetup{
  colorlinks=false,
  linkbordercolor=white,
 urlbordercolor=white,
pdfborder={0 0 0}
}

\newcolumntype{C}[1]{>{\centering\arraybackslash}p{#1}}

\newcommand{\fheon}{{\textsc{\small{FHEON}}}\xspace}
\newcommand{\openfhe}{{\textsc{\small{OpenFHE}}}\xspace}

\DeclareMathOperator{\Rot}{Rot}

\DeclareMathOperator{\Enc}{Enc}
\DeclareMathOperator{\Sum}{Sum}
\DeclareMathOperator{\Merge}{Merge}

\begin{document}

\title{\Large\bfseries Towards Deep Encrypted Training: Low-Latency, Memory-Efficient, and High-Throughput Inference for Privacy-Preserving Neural Networks}

\author{
Nges Brian Njungle \orcidlink{0009-0006-3393-6851}, Eric Jahns \orcidlink{0009-0004-5511-7975}, and Michel A. Kinsy  \orcidlink{0000-0002-1432-6939} \\
STAM Center, Ira A. Fulton Schools of Engineering\\
Arizona State University, Tempe, AZ 85281, USA\\
Emails: nnjungle@asu.edu, jjahns@asu.edu, mkinsy@asu.edu
}

\maketitle

\input{sections/abstract}

\input{sections/Introduction}
\input{sections/Related_Work}
\input{sections/Background}
\input{sections/Threat_Model}
\input{sections/Methodology}

\input{sections/Pipeline}
\input{sections/Experiments}
\input{sections/Results}
\input{sections/Conclusion}
\bibliographystyle{plain}
\bibliography{paper}

\end{document}

%% file: sections/Abstract.tex
\begin{abstract}
Privacy-preserving machine learning (PPML) has become increasingly important in applications where sensitive data must remain confidential. Homomorphic Encryption (HE) enables computation directly on encrypted data, allowing neural network inference without revealing raw inputs. While prior works have largely focused on inference over a single encrypted image, batch processing of encrypted inputs lags behind, despite being critical for high-throughput inference scenarios and training-oriented workloads.

In this work, we address this gap by developing optimized algorithms for batched HE-friendly neural networks. 
We also introduced a pipeline architecture designed to maximize resource efficiency for different batch size execution. We implemented these algorithms
and evaluated our work using HE-friendly ResNet-20 and ResNet-34 models on encrypted CIFAR-10 and CIFAR-100 datasets, respectively.
For ResNet-20, our approach achieves an amortized inference time of $8.86$ seconds per image when processing a batch of $512$ encrypted images, with a peak memory usage of $98.96 $~GB. 
These results represent a $1.78\times$ runtime improvement and a $3.74\times$ reduction in memory usage compared to the state-of-the-art design. 
For the deeper ResNet-34 model, we achieve an amortized inference time of $28.14$ 
on a batch of $256$ encrypted images using $246.78$~GB of RAM.
\end{abstract}

%% file: sections/Introduction.tex
\section{Introduction}
\label{sec:intro}

Over the past decade, Machine Learning (ML) has evolved into a cornerstone of modern technology, driving innovation across nearly every sector of society~\cite{angra2017machine}. 
From personalized healthcare and finance to autonomous driving systems and cybersecurity, ML models now underpin a wide range of data-driven applications. 
Advances in model architectures, computing hardware, and large-scale data availability have further propelled ML toward solving increasingly complex real-world problems, making it a critical enabler of intelligent automation and decision-making across many  applications~\cite{sarker2023machine}.  

One fundamental paradigm that has accelerated the accessibility and scalability of ML is Machine Learning as a Service (MLaaS)~\cite{ribeiro2015mlaas}.
MLaaS enables model developers and organizations to deploy highly efficient and complex models through cloud-based infrastructure, providing inference capabilities as a utility. 
Here, users can remotely access pre-trained models, submit queries, and receive predictions without maintaining local models. 
Another important direction in modern ML is collaborative training~\cite{tseng2024co}.
Here, multiple entities jointly train a model to leverage distributed data resources and domain expertise. 
This approach allows organizations to train a more intelligent and generalized model without explicitly sharing their datasets. 
A more recent, but increasingly relevant domain is cross-organizational model evaluation~\cite{yang2024intelligent}. 
In this setting, one party develops a sophisticated model, while another party holds the data needed to test or fine-tune that model’s performance. 
Cross-organization evaluation enables stakeholders to assess the utility of external models without requiring full data or model transfer, thus promoting collaborative innovation and interoperability between independent entities.  

Although these domains differ in structure and objective, they all share a common challenge when handling sensitive data. 
Whether in MLaaS deployment, collaborative training, or cross-organization evaluation, the exchange of sensitive model weights and sensitive user data introduces critical privacy and security risks. 
These risks become especially acute when ML systems process sensitive or personally identifiable data such as genomic sequences, medical records, or financial transactions~\cite{papernot2018sok}. 
Yet, these are also the very domains that stand to benefit the most from the efficiency, scalability, and intelligence of modern ML systems. 
This tension between utility and confidentiality has made privacy-preserving machine learning a central focus for many recent ML works.  

Privacy-preserving machine learning (PPML) seeks to enable the benefits of ML while ensuring that sensitive information remains confidential throughout the computation process~\cite{mohassel2017secureml}. In the application domains discussed above, PPML techniques allow multiple parties to train, inference, or deploy models without exposing raw data or revealing proprietary model parameters.
Among the various approaches used for PPML today, Homomorphic Encryption (HE) has emerged as a particularly compelling solution. HE allows computations to be performed directly on encrypted data, producing encrypted outputs that can be decrypted to obtain the same results as computations on plaintext inputs~\cite{acar2018survey}. This property enables ML systems to perform inference and training without ever exposing sensitive information. 
Despite their promise, today's practical HE schemes often impose substantial computational and memory overheads in applications, making encrypted neural network models orders of magnitude slower than their plaintext counterparts. Addressing these performance bottlenecks is essential to enable PPML with HE at scale, particularly in deep learning models~\cite{gouert2023sok, njungle2025guardianml}.

\subsection{Limitations of Encrypted Neural Networks}

Within the context of HE-based neural networks, a primary approach is single-input processing. Here, each input is encrypted and processed individually through a HE-friendly neural network model~\cite{rovida_cnn}. 
This method is particularly valuable in scenarios that demand real-time processing of sensitive queries, such as personalized medical diagnosis, confidential financial decision support, or secure recommendations for individual users. 
In fact, the most advanced encrypted neural networks of today are focused on single-input processing. 
Recent frameworks such as FHEON~\cite{njungle2025fheon} and Orion~\cite{ebel2025orion} have demonstrated stable implementations of encrypted deep neural networks and have significantly pushed the performance limits of HE inference in this setting.
Despite its importance and numerous applications, this single-input processing approach is highly limited in high-throughput environments, as each input requires a complete sequence of computationally expensive homomorphic operations. 

The second approach to encrypted neural networks is encrypted batch input processing. 
In this method, multiple encrypted inputs are processed simultaneously through a HE-friendly neural network model~\cite{cheon2024batch}. 
Encrypted batch image processing can greatly improve throughput and reduce amortized latency compared to encrypted single-input processing. 
However, this approach is highly constrained by the complexity of packing multiple images, complexity of model design, the number of HE operations allowed within a ciphertext, and the substantial memory and computational requirements of processing these large amounts of encrypted data. 

The importance of encrypted batch inference becomes particularly evident in real-world applications that require processing large volumes of sensitive data at once. For example, hospitals may need to run predictions over hundreds of encrypted scans simultaneously to support clinical decision-making. Financial institutions may wish to evaluate large batches of transaction data for fraud detection or risk assessment without exposing customer information.

Beyond inference, achieving efficient encrypted batch processing is also a prerequisite for practical HE-based deep neural network training. 
Training deep neural networks relies on batch operations for efficient gradient computation and weight updates, since models generally require large amounts of data to learn and generalize effectively~\cite{kandel2020effect}.

Many previous works have proposed different approaches for encrypted batch input processing. The state-of-the-art work of Cheon et al. introduced the most optimized approach for this line of work~\cite{cheon2024batch}.
Their implementations leveraged multi-threading on a server-grade computational setup, resulting in an encrypted-image amortized runtime of about $15.76$ seconds in a 512-image pipeline using a HE-friendly ResNet-20 model and the CIFAR-10 dataset.
While their work showed about $5\times$ performance improvement over the previous version, it still requires a memory footprint of $370.28$~GB of RAM which is still very high. 
Improving encrypted batch processing is therefore critical not only for high-throughput encrypted inference workloads but also as a major step towards practical HE-based deep neural networks training.

\subsection{Contributions of this Work}

In this work, we take an in-depth look at batch processing of encrypted images using HE-friendly deep neural networks. 
We then develop novel and highly efficient algorithms required for the different neural network layers, focusing on reducing both the memory footprint and latency. 
Improving the individual layers is important as they are transferable across different architectures, enabling the effective evaluation of a wide range of deep models. 

To further improve throughput, we propose an optimized pipeline design that keeps encrypted neural network execution continuously active by maximizing HE resource utilization. Our approach introduces an accumulator that merges intermediate ciphertexts across different neural network groups, ensuring that the pipeline remains saturated and avoids underutilization.
Beyond sustaining high utilization, the accumulator provides a principled mechanism for controlling the effective batch size during encrypted inference. This capability is critical, as batch size directly impacts the trade-off between latency and throughput. Smaller batch sizes reduce per-query latency, while larger batch sizes improve amortized latency and overall throughput.
By enabling flexible batch size control within a unified execution pipeline, the accumulator also allows our design to adapt to varying application requirements, thereby improving both system efficiency and practical deployability.

To evaluate our work, we developed our proposed high-throughput neural network layers on top of \fheon~\cite{njungle2025fheon} and the OpenFHE library~\cite{OpenFHE}. 
\fheon provides an open-source configurable PPML inference framework developed using OpenFHE's implementation of the Cheon-Kim-Kim-Song scheme~\cite{ckks}. 
It provides robust and extensible encrypted neural network layers for single-input inference, making it an ideal foundation for this work. 
From our layers, we implemented high-throughput HE-friendly ResNet-20 and ResNet-34 models and evaluated them on the CIFAR-10 and CIFAR-100 datasets, respectively. 
Our evaluation spans batch sizes ranging from 16 to 512 encrypted images under different encryption parameter settings. 
We achieved an amortized inference time of $8.86$ seconds per encrypted image for a 512-image batch while requiring just 98.96~GB of RAM. 
On the other hand, our ResNet-34 model achieved an amortized time of $28.14$ seconds per encrypted image with a batch size of 256, while requiring 246.78~GB of RAM. 
Concretely, this work makes several key contributions:
\begin{itemize}

   \item We develop optimized algorithms that improve ciphertext packing efficiency, reduce unnecessary rotations, and streamline layer-level operations to support large-scale encrypted batch-input neural network inference.
       
    \item We propose a pipeline architecture that maximizes models's throughput by keeping network layers continuously active. This design minimizes idle computation and empty ciphertext slots, enabling efficient data flow across all stages of encrypted execution.
    
    \item We evaluate our work using twelve HE-friendly ResNet-20 and ResNet-34 models on the CIFAR-10 and CIFAR-100 datasets, respectively, and across batch sizes ranging from 16 to 512 images. This extensive evaluation demonstrates the scalability, efficiency, and practical viability of our methods for real-world PPML application scenarios.

\end{itemize}

%% file: sections/Related_Work.tex
\section{Related Works}
\label{sec:related_works}

Research on HE-based neural networks has evolved across several generations of HE schemes, optimizations, and system designs. 
The earliest practical demonstration came from Microsoft Research through CryptoNets~\cite{gilad2016cryptonets} in 2016, which showed that simple neural network models can be evaluated in the encrypted domain. 
This initial work catalyzed a large body of encrypted neural networks exploring different HE schemes, packing strategies, activation functions, convolution strategies, and bootstrapping procedures.

Early systems such as Faster CryptoNets~\cite{chou2018faster}, Brutzkus et al.~\cite{brutzkus2019low}, Falcon~\cite{lou2020falcon}, and HCNN~\cite{al2020towards} primarily relied on the BGV~\cite{bgv} and BFV~\cite{bfv} schemes. 
These HE schemes encode multiple integers into a Polynomial using a large modulus encoded with the Chinese Remainder Theorem, enabling Sing instruction mutiple data(SIMD) style vectorized encrypted computations. To operate within these integer-based schemes, these works employed quantization techniques to map floating-point neural network weights into the supported integer domain. 
As a result of the high cost of bootstrapping and the limitations associated with using polynomial activation functions such as the square function, they were restricted to shallow network depths and achieved limited accuracies.

A recent line of research have explored TFHE~\cite{tfhe} for encrypted neural network inference. TFHE is a bit-wise HE scheme that supports arbitrary boolean circuits hence exact implementations of nonlinear operations such as ReLU and maxpooling. 
TFHE-based approaches can therefore yield very high accurate encrypted neural networks, but they do not support SIMD computations thus computationally expensive for deep models. 
These works require weight discretizations further increasing model size and runtime. The state-of-the-art work in this line of research was published by  Benamira et al.~\cite{benamira2023tt}. The authors evaluated encrypted CIFAR-10 images on a few-layer custom CNN in 2256 seconds per image.

The most widely adopted approach for encrypted neural network evaluation is based on the CKKS scheme~\cite{ckks}. CKKS is an approximate HE scheme that natively supports real-number arithmetic and SIMD-style parallelism. 
Its ability to efficiently evaluate polynomials with  high-precision has led to it's adoption and the use of approximations of nonlinear activation functions. These advantages together with comparatively efficient bootstrapping techniques, has made CKKS particularly well suited for deep encrypted neural networks. 

Existing CKKS-based deep encrypted neural network systems can be broadly categorized into single-input and high-throughput inference works.

\subsection{Single-Input Works}

Most of the literature on encrypted neural networks today fall under this caterogry. 
Representative works have been benchmarked on a HE-friendly ResNet-20 model and encrypted CIFAR-10 dataset include; Lee et al. (2022), who report 91.31\% accuracy with an inference latency of 2,271 seconds \cite{lee2022privacy}. Kim et al. (2023), pushed the inference latency to just 255 seconds~\cite{kim2023optimized}. Rovida et al. (2024), further reduced the memory requirements of encrypted inference to just 15.1 GB of RAM \cite{rovida_cnn} with an inference time of 331 seconds. 

Beyond standalone optimizations, more recent research has moved towards the production of comprehensive frameworks that make the development of single-input HE-friendly neural networks more accessible and friendly. 
Two notable systems in this space are FHEON (2025)~\cite{njungle2025fheon} and Orion (2025)~\cite{ebel2025orion}. FHEON is an open-source configurable framework that exposes HE-friendly CNN layers and optimized primitives (convolution, average pooling, ReLU approximations, and fully connected layers) and supports arbitrary CNN architectures through parameterized layer implementations. 
Orion on the other hand is a compiler-style FHE framework that accepts PyTorch-like models and translates them into efficient equivalent HE programs. It also apply a set of predefined optimizations to these derived HE-friendly models.

\subsection{High-Throughput Works}

 Works in this category targets batch inference, aiming to reduce the high encrypted inference time per input by evaluating multiple inputs through a HE-friendly neural network. 
Early efforts such as~\cite{gilad2016cryptonets, nandakumar2019towards, xu2019cryptonn, boemer2019ngraph} were limited to small-scale networks evaluated on simple datasets like MNIST.
They mostly achieved batching through coarse-grained parallelism (multi-threading or multi-process orchestration of single-image executions), which increases throughput but scales very poorly~\cite{gilad2016cryptonets}. 
More recent algorithmic approaches have focused on packing multiple images into ciphertext slots  such as those introduced by Park et al.~\cite{park2023toward}, Cheon et al.~\cite{cheon2024batch}, HTCNN~\cite{11223652}.
The most efficient of these works was published in 2024 by Cheon et al~\cite{cheon2024batch} as a  5.04$\times$ improvement over their 2023 version~\cite{park2023toward}. 
The authors introduced a channel-by-channel convolution packing approach that places different channels of multiple images into separate ciphertexts.
The work achieved  a low amortized latency of 15.76 seconds per encrypted image with a memory requirement of  370.28 GB of RAM using a ResNet-20 model with CIFAR-10 images. 

Over the past decade of encrypted neural network research (2016--2026), substantial progress has been achieved in activation function approximation, ciphertext packing strategies, compiler-level automation, and now the development of frameworks that provide production-quality encrypted deep neural network primitives. 
Despite these advances, batched encrypted neural network inference remains significantly less mature than single-input inference.
In this work, we push high-throughput encrypted neural network inference to a level also suitable for practical deployment. 
We introduce novel algorithmic techniques that jointly reduce inference latency and memory consumption for batched encrypted inference. 
Our design targets a realistic mid-range server configuration with 32 CPU cores and 256~GB of RAM, avoiding the reliance on high-end, specialized hardware commonly assumed in prior works. 
Moreover, our approach reduces the memory footprint of most batched encrypted inference scenarios to below 128~GB of RAM, substantially improving accessibility and deployability.
We implement our proposed techniques in \fheon and demonstrate significant performance gains. Compared to the most advanced batched encrypted inference design of Cheon et al.~\cite{cheon2024batch}, our system achieves approximately a $2\times$ reduction in runtime and a $4\times$ reduction in memory usage, while operating under practical hardware assumptions.

%% file: sections/Background.tex
\section{Background}
\label{sec:background}

\subsection{Homomorphic Encryption}

Homomorphic Encryption (HE) is a cryptographic technique that enables computations to be carried out directly on encrypted data. 
In HE, operations performed on ciphertexts produce encrypted outputs which match the results of the same operations performed in plaintext once decrypted~\cite{naehrig2011can}. This property allows untrusted servers to compute on sensitive data without ever learning the underlying values.
Fully Homomorphic Encryption (FHE) was first demonstrated by Gentry in 2009 \cite{gentry}. 
Gentry introduced the idea of bootstrapping, which refreshes a ciphertext by homomorphically evaluating the decryption circuit to reduce accumulated noise. 
Bootstrapping enables arbitrarily encrypted computations by periodically restoring the noise budget required for correctness and security, thereby extending HE to fully homomorphic capability.

\subsection{The CKKS Scheme}

The Cheon–Kim–Kim–Song (CKKS) scheme is an approximate HE scheme designed to support computation on real or complex-valued data~\cite{ckks}. Its security is based on the hardness of the Ring Learning with Errors (RLWE) problem, which extends the standard Learning with Errors problem \cite{lwepropsed, ringlwe} to polynomial rings.
RLWE provides strong security guarantees while enabling efficient algebraic operations over polynomial structures.
CKKS encodes vectors of complex numbers into polynomials in the cyclotomic ring shown in Equation~\ref{eq:rlwe} where \(N\) is a power of two.
\begin{equation}
    \label{eq:rlwe}
      R = \mathbb{Z}[X] / (X^N + 1),
\end{equation}
 Through canonical embeddings, CKKS maps a vector \(v \in \mathbb{C}^{N/2}\) into a polynomial in \(R\), allowing up to \(N/2\) independent data slots in a single ciphertext. 
 
CKKS supports homomorphic addition, multiplication, rotation, and bootstrapping operations.
The practical performance of CKKS-based neural network inference depends on several core mechanisms:

\paragraph{Ciphertext Packing and SIMD Parallelism.}
Packing many values into a single ciphertext allows large parallelism and amortizes computation. Convolution layers in particular benefit from packing strategies that minimize the number of ciphertexts required to encode feature maps or channels. 
However, packing CKKS is constrained by the need for conjugate symmetry in real-values encoding, which limits the number of independent slots to \(N/2\).

\paragraph{Rotations and Key Switching.}
Rotations are required for implementing convolutions and matrix–vector multiplications of the linear layers. Each rotation triggers a key-switch operation, which is often the dominant cost in CKKS execution. Efficient key-switching algorithms and precomputed rotation keys significantly reduce runtime, but rotations still represent more than half the total inference cost.

\paragraph{RNS Decomposition and NTT Optimization.}
Modern CKKS implementations are based on the Residual Number Systems variant (RNS-CKKS) and Number Theoretic Transforms (NTT) which accelerates the underling polynomial arithmetic~\cite{rns_ckks}. 
RNS decomposition allows large integers to be represented as vectors of smaller moduli, enabling faster operations and parallel execution. NTT-based polynomial multiplication reduces the complexity of ring multiplication from quadratic to quasi-linear. 
The faster the operations, the better it is for the deep networks.

\paragraph{Bootstrapping.}
Bootstrapping restores the noise budget and allows evaluation of deeper networks. 
Although it has been significantly optimized over the years, bootstrapping remains expensive and motivates architectural designs that minimize the number of bootstraps required.

\paragraph{Memory Requirements.}
CKKS ciphertexts are large objects consisting of multiple polynomials. 
Bootstrapping and rotation keys are also very large objects required throughout the HE-friendly models for different computations. 
Model parameters must be encoded into ciphertexts or plaintext polynomials which also consume substantial memory. 
Packing strategies, amount of keys required, and neural network layers designs heavily influence the memory requirements of encrypted networks. 
Thus, memory requirement is a major bottleneck in most encrypted neural networks designs today.

\subsection{FHEON and OpenFHE}

This work is build on \fheon~\cite{njungle2025fheon} which is intend build on the CKKS implementation provided by the OpenFHE library \cite{OpenFHE}. 
OpenFHE implements the RNS-CKKS scheme introduced in \cite{rns_ckks}. 
The library also includes optimized support for ciphertext packing, fast rotations \cite{fast_rotations}, advanced key-switching, and an efficient bootstrapping engine. 
It also exposes APIs for parallel execution accelerated vector operations~\cite{boemer2021intel}. 
\fheon leverages OpenFHE to develop configurable single-input inference layers as well as utility functions that support the development of encrypted neural networks. It also provide an update documentation and it is actively maintained thus, an ideal choice for this work. 

%% file: sections/Threat_Model.tex
\section{Threat Model}

This work adopts the same threat model adopted across most prior encrypted neural networks. 
The system involves two parties: a client and a server. 
The client is assumed to be \emph{honest}, meaning it follows the protocol exactly and does not attempt to infer information beyond the prescribed output. 
The server, which performs inference on encrypted data, is assumed to be \emph{semi-honest} (honest-but-curious). That is, the server correctly executes the protocol as specified, but may attempt to learn additional information about the client’s private data by observing encrypted inputs, intermediate ciphertexts, or execution behavior.

In this setting, the server hosts a trained neural network model and retains the model parameters in plaintext.
This reflects a typical out-sourced encrypted neural network inference deployment where the service provider owns and manages the model. 
The client encrypts its input data locally using the CKKS scheme and transmits the ciphertext to the server. 
The server performs inference directly on these encrypted inputs using its plaintext model parameters and returns an encrypted prediction to the client. 
Since the secret decryption key is known only to the client, the server cannot decrypt either the client’s inputs or the inference result, and thus learns no information about the client’s data.

This threat model provides strong confidentiality guarantees for client inputs while allowing the server to deploy proprietary models without exposing them to the client. 
Although the system can be extended to support encrypted model weights for models confidentiality, we do not consider this setting in the present setup.
Also consistent with most HE-based inference systems, our work does not address adversaries that actively deviate from the protocol, nor does it consider side-channel attacks such as timing, cache, or power analysis. Availability attacks on the server such as denial-of-service, are also outside the scope of this threat model.

%% file: sections/Methodology.tex
\section{Methodology}

We begin by examining the different layers of neural networks and then transformed them for efficient execution in the encrypted domain. 
Our designs also minimize data dependencies in order to effectively leverage multi-threaded CPU architectures. 
Based on these principles, we develop optimized designs for batch input representations, batch convolution operations, and batch fully connected layers.

\subsection{\textbf{Encrypted Input Representation}}

The work of Cheon et al.~\cite{cheon2024batch} introduced an optimized approach for representing batch encrypted inputs for encrypted batch neural network processing called Channel-By-Channel (CBC) packing.  
Here, the channels of every input are encoded across multiple input ciphertexts. 
Each ciphertext encodes the same channel for multiple inputs align one after the other. 
We adopt this proposed approach for processing inputs and extended \fheon to handle it as well as dynamic weights encoding for all neural network layers required in the processing of these types of encrypted inputs.  
This layout enables efficient channel-wise convolution while preserving batch-level parallelism

Let $n$ denote the number of available slots in a ciphertext and let $b$ denote the batch size. Consider an input tensor with $p$ channels, where each channel contains $m$ values corresponding to a flattened spatial dimension. 
The input is represented as a vector of it's flattened channels as shown in Equation~\ref{eq:input-ciphertexts} where each $C_i$ encodes the $i$-th channel of the input.
\begin{equation}
\mathbf{C} = \left( C_1, C_2, \ldots, C_p \right),
\label{eq:input-ciphertexts}
\end{equation}

For CBC encoding, we rearrange the plaintext of the various inputs into SIMD vectors and encrypted them into multiple ciphertext $C_i$ as shown in~\ref{eq:cbc-packing}. $\mathbf{x}_{i}^{(j)} \in \mathbb{R}^m$ denotes the flattened spatial values of channel $i$ for the $j$-th input sample. 
\begin{equation}
C_i = \Enc\left(
\begin{bmatrix}
\mathbf{x}_{i}^{(1)} \mid \mathbf{x}_{i}^{(2)} \mid \cdots \mid \mathbf{x}_{i}^{(b)}
\end{bmatrix}
\right),
\label{eq:cbc-packing}
\end{equation}

To avoid slot overlapping and to ensure correct alignment across batch segments, the constraint Equation shown Equation~\ref{eq:slot-constraint} must hold.

\begin{equation}
m \cdot b \leq n.
\label{eq:slot-constraint}
\end{equation}
This constraint directly governs the maximum achievable batch size for a given ciphertext polynomial degree.

In addition, the packed layout must preserve contiguity of spatial data within each batch segment. Specifically, for the $j$-th input in the batch, the channel data vector $\mathbf{x}_i^{(j)}$ must occupy a contiguous block of $m$ slots, starting at index $j \cdot m$. Formally, the slot indices assigned to $\mathbf{x}_i^{(j)}$ are given by by Equation~\ref{eq:slot-alignment}.
\begin{equation}
\mathbf{x}_i^{(j)} \;\mapsto\; \left\{ j \cdot m,\; j \cdot m + 1,\; \ldots,\; (j+1)\cdot m - 1 \right\}.
\label{eq:slot-alignment}
\end{equation}

\begin{figure}[ht]
    \centering
    \includegraphics[width=\linewidth]{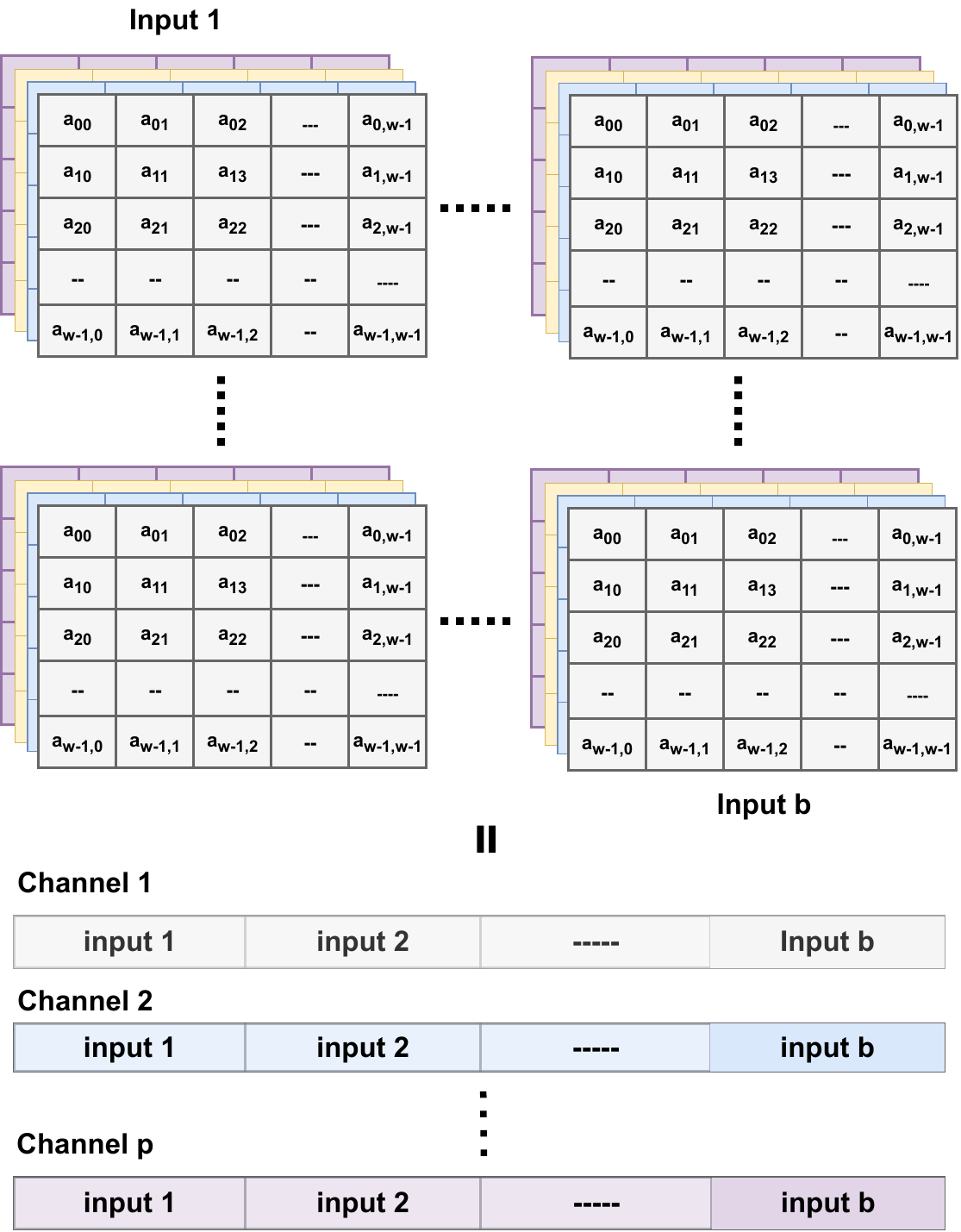}
    \captionsetup{justification=centering}
    \caption{Transforming multiple inputs into multiple SIMD vectors using channel-by-channel encoding.}
    \label{fig:input_encoding}
\end{figure}

Figure~\ref{fig:input_encoding} shows how multiple inputs  are transformed and encoded into multple input ciphertexts. These ciphertexts are then passed into the HE-friendly neural network for encrypted inference.  

\subsubsection{Weights Encoding}

Let the convolution kernel consist of $p$ input channels, each with a spatial support of size $t$. The kernel weights are generally represented as shown in Equation~\ref{eq:kernel-def} where $\mathbf{w}_i = (w_{i,0}, \ldots, w_{i,t-1}) \in \mathbb{R}^t$.
\begin{equation}
\mathbf{W} = \left( \mathbf{w}_1, \mathbf{w}_2, \ldots, \mathbf{w}_p \right),
\label{eq:kernel-def}
\end{equation}

To enable efficient convolutions, we encode each kernel vector $\mathbf{w}_i$ into a plaintext vector $\mathbf{w}_i^{\text{enc}} \in \mathbb{R}^n$ by dynamically replicating the kernel weights across the entire batch layout as shown in Equation~\ref{eq:kernel-packing}.
\begin{equation}
\mathbf{w}_i^{\text{enc}} =
\begin{bmatrix}
\mathbf{w}_i^{(1)}  \mid \mathbf{w}_{i}^{(2)} \mid \cdots \mid \mathbf{w}_i^{(b)} 
\end{bmatrix},
\label{eq:kernel-packing}
\end{equation}
where each block corresponds to a single input in the batch and is aligned with the spatial positions of the packed channel data. 
This encoding enables a single plaintext-ciphertext multiplication to apply the same kernel weights simultaneously across the entire batch.

\input{sections/_convolutionlayer}

\input{sections/_fclayer}

\begin{figure*}[http]
    \centering
    \includegraphics[width=0.78\textwidth]{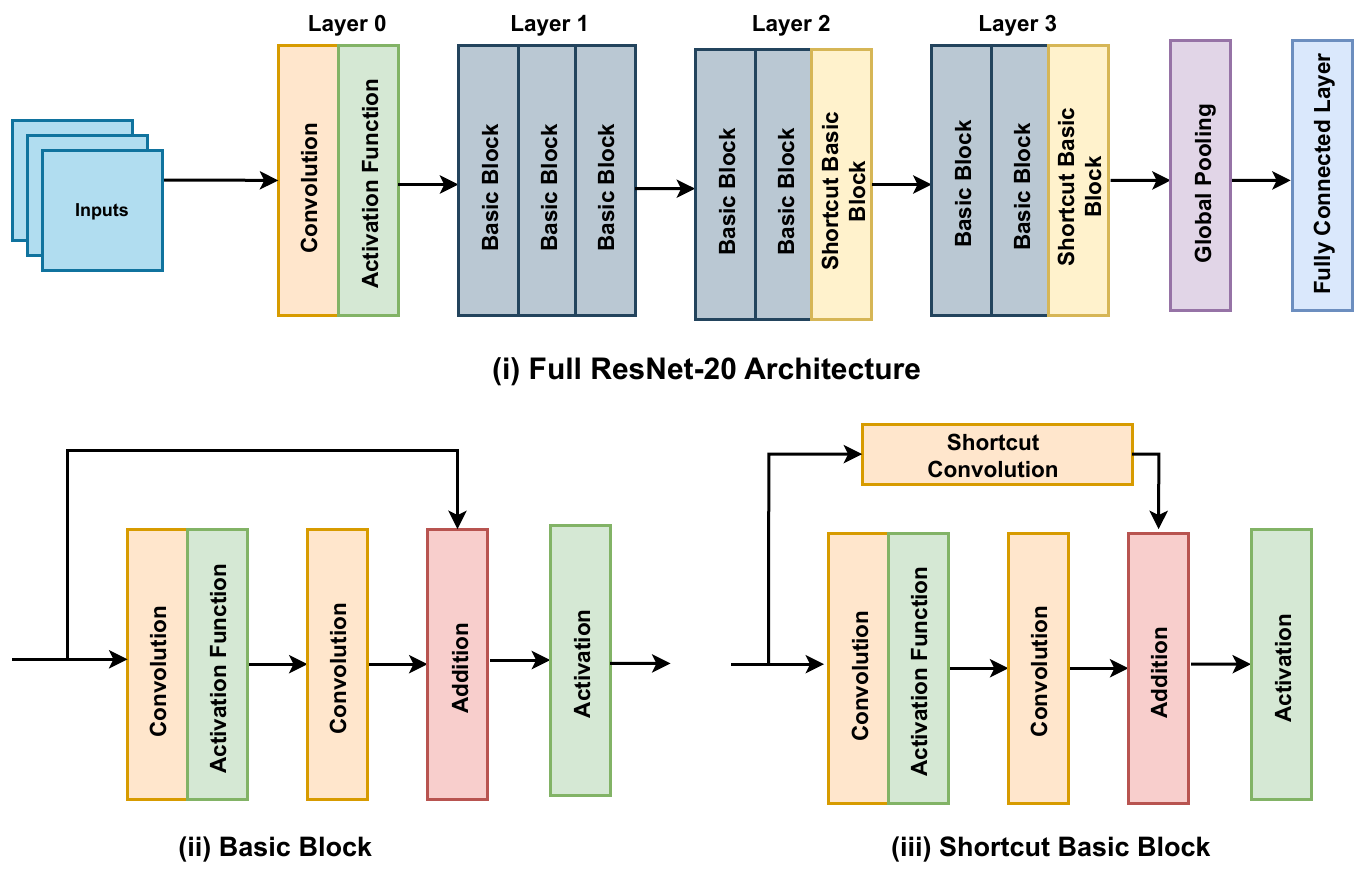}
    \captionsetup{justification=centering}
    \caption{The standard ResNet-20 architecture used as the baseline network in this work. The model consists of an initial convolution layer followed by three residual stages with increasing channel widths and periodic downsampling. The Basic Block and the Shortcut Basic Block sub modules shows their inner structures }
    \label{fig:basic_pipeline}
\end{figure*}

\input{sections/_multithreading}

\input{sections/_otherlayers}

%% file: sections/_convolutionlayer.tex
\subsection{Convolution Layer}
\label{sec:batch-convolution}

A convolution layer applies a set of kernels to the input data by sliding each kernel across the spatial dimensions of the input. At each spatial location, an element-wise multiplication is performed between the kernel weights and the corresponding input region, followed by a summation to produce a single output value~\cite{Kamath2019}. 
When evaluated under HE, this process must be carefully reformulated to minimize the number of expensive HE operations needed. 

To enable homomorphic convolution, the encrypted input must first be aligned such that the kernel weights can be applied using plaintext-ciphertext multiplications. 
A naive approach requires rotating the input ciphertext by all $k^2$ spatial offsets for a $k \times k$ kernel, resulting in $k^2 -1$ rotation keys.
Instead, we introduce an optimized rotation strategy for batch encrypted input processing that reduces the number of required rotation keys for every convolution layer to just four keys. This approach relies on a small set of base rotations, which are then reused to derive all required spatial alignments through additional rotations. The key idea is to generate intermediate ciphertexts using a limited set of fast rotations, and subsequently apply inexpensive re-rotations to obtain the full set of rotated ciphertexts required for convolution.
This type of approach was originally proposed for single-input convolution in Rovida et al.~\cite{rovida_cnn} and our work extends it to batched encrypted inputs.

\subsubsection{Input Alignment via Optimized Rotations}
\label{sub:input_alignment}

Let $\texttt{ciInput}$ denote the ciphertext encoding a single input channel. We first generate two intermediate ciphertexts from the input by applying unit rotations in opposite directions, as shown in Equations~\ref{eq:base-rotations_1},~\ref{eq:base-rotations_2}.
\begin{equation}
C^{-1} = \Rot(\texttt{ciInput}, -1) 
\label{eq:base-rotations_1}
\end{equation}
\begin{equation}\
C^{+1} = \Rot(\texttt{ciInput}, 1).
\label{eq:base-rotations_2}
\end{equation}

Using these base ciphertexts, we derive the remaining spatial alignments by applying vertical rotations of magnitude $\pm W$, where $W$ denotes the input width to the three base ciphertexts.
These rotations then create the $k \times k$ ciphertexts required for the convolution window. In the case of $3 \times 3$ kernel window,  we generate the set of ciphertexts shown in Equation~\ref{eq:rotation-set}.
Each ciphertext in $\mathcal{R}$ corresponds to one spatial offset of the convolution kernel relative to the input. 
Importantly, all required rotations are derived from only four rotation keys: $\{-1, +1, -W, +W\}$, drastically reducing the memory requirements of storing nine rotation keys.

\begin{equation}
\begin{aligned}
\mathcal{R} = \{ \;
& \Rot(C^{-1}, -W), \\
& \Rot(\texttt{ciInput}, -W), \\
& \Rot(C^{+1}, -W), \\
& C^{-1}, \\
& \texttt{ciInput}, \\
& C^{+1}, \\
& \Rot(C^{-1}, W), \\
& \Rot(\texttt{ciInput}, W), \\
& \Rot(C^{+1}, W)
\; \}.
\end{aligned}
\label{eq:rotation-set}
\end{equation}

\subsubsection{Convolution Evaluation}
\label{sub:convolution_evaluation}

We evaluate the convolution operation by multiplying each aligned ciphertext with its corresponding encoded kernel weight and accumulating the results as shown in Equation~\ref{eq:homomorphic-conv}.
\begin{equation}
C_{\text{conv}} = \sum_{\ell=1}^{k^2} \mathcal{R}_\ell \odot w_\ell,
\label{eq:homomorphic-conv}
\end{equation}
where $\mathcal{R}_\ell$ denotes the $\ell$-th rotated ciphertext and $w_\ell$ is the corresponding encoded kernel weight.

A key advantage of our design is that it completely eliminates the additional rotation operations that are required to realign intermediate results across channels prior to summation. 
With the CBC packing, each ciphertext is made up of data for a channel corresponding across the entire batch. 
As a result, channel-wise accumulation reduces to direct homomorphic additions of the ciphertexts, requiring no further rotations.
This procedure is repeated independently for each output channel, resulting in a vector of output ciphertexts that collectively represent the encrypted output feature maps.

The elimination of intermediate rotations over input channels further reduces the number of required rotation keys, which further lowers the overall memory footprint, while preserving correctness and full compatibility with batched ciphertext layouts. 
Furthermore, since the same rotation pattern is applied uniformly across all packed inputs, the convolution layer is evaluated simultaneously for all inputs in the batch.

%% file: sections/_fclayer.tex
\subsection{Fully Connected Layer}
\label{sec:fully-connected}

The fully connected (FC) layer, also referred to as a linear layer is geernally used in the neural networks for classification. It performs an affine transformation of an input as shown in Equation~\ref{eq:fc-definition} 
where $\mathbf{x} \in \mathbb{R}^{d_{\text{in}}}$ denotes the input vector, $\mathbf{W} \in \mathbb{R}^{d_{\text{out}} \times d_{\text{in}}}$ is the weight matrix, and $\mathbf{b} \in \mathbb{R}^{d_{\text{out}}}$ is the bias vector.
\begin{equation}
\mathbf{y} = \mathbf{W}\mathbf{x} + \mathbf{b},
\label{eq:fc-definition}
\end{equation}

In the homomorphic setting, direct matrix-vector multiplication is not supported. 
Instead, the FC layer must be reformulated using ciphertext rotations, plaintext-ciphertext multiplications, and slot-wise summations. In this work, we proposed a batched FC layer that evaluates Equation~\ref{eq:fc-definition} simultaneously over multiple inputs packed into a single ciphertext.

\subsubsection{FC Evaluation}
\label{sub:fc-evaluation}
Let $b$ denote the batch size, $d_{\text{in}}$ the input dimension, and $d_{\text{out}}$ the output dimension. The encrypted input ciphertext $C_{\text{in}}$ encodes a vector of inputs as shown in Equation~\ref{eq:fc-input-packing} where each $\mathbf{x}^{(j)} \in \mathbb{R}^{d_{\text{in}}}$ occupies a contiguous block of $d_{\text{in}}$ slots. 
\begin{equation}
C_{\text{in}} = \Enc\left(
\mathbf{x}^{(0)} \mid \mathbf{x}^{(1)} \mid \cdots \mid \mathbf{x}^{(b-1)}
\right),
\label{eq:fc-input-packing}
\end{equation}

Correctness requires that the constraint of Equation~\ref{eq:fc-slot-constraint} is satisfied where $n$ is the ciphertext slot count.
\begin{equation}
b \cdot d_{\text{in}} \leq n,
\label{eq:fc-slot-constraint}
\end{equation}

Each row of the weight matrix $\mathbf{W}$ is encoded as a plaintext vector aligned with the input layout. For output index $i \in \{0, \ldots, d_{\text{out}}-1\}$, the weight vector $\mathbf{w}_i \in \mathbb{R}^{d_{\text{in}}}$ is replicated across the batch layout as shown in Equation~\ref{eq:fc-weight-encoding}
\begin{equation}
\mathbf{w}_i^{\text{enc}} =
\begin{bmatrix}
\mathbf{w}_i^{(1)}  \mid \mathbf{w}_{i}^{(2)} \mid \cdots \mid \mathbf{w}_i^{(b)} 
\end{bmatrix}.
\label{eq:fc-weight-encoding}
\end{equation}

We then compute an element multiplication as shown in Equation~\ref{eq:fc-multiplication}.
To compute the inner product for each output channel in the batch, we perform a slot-wise summation over contiguous blocks of input data in the batch as shown in Equation~\ref{eq:fc-inner-sum} where $\Sum_{d_{\text{in}}}(\cdot)$ denotes summation over $d_{\text{in}}$ slots and $\Rot(\cdot)$ is the ciphertext rotation operator.
\begin{equation}
C_i = C_{\text{in}} \odot \mathbf{w}_i^{\text{enc}}
\label{eq:fc-multiplication}
\end{equation}
\begin{equation}
\widetilde{C}_i^{(j)} = \Sum_{d_{\text{in}}}\!\left(
\Rot\!\left(C_i, j \cdot d_{\text{in}}\right)
\right),
\label{eq:fc-inner-sum}
\end{equation}

\subsubsection{Output Reconstruction}
\label{sub:output-reconstruction}

To efficiently reconstruct the batched output layout, for each batch index $j$,   ciphertexts containing the results for the corresponding output channels are merged using ciphertext concatenation as shown in Equation~\ref{eq:fc-merge}.
\begin{equation}
C_{\text{out}}^{(j)} = \Merge\!\left(
\widetilde{C}_0^{(j)}, \widetilde{C}_1^{(j)}, \ldots
\right).
\label{eq:fc-merge}
\end{equation}

The merged ciphertexts are then rotated to their global slot positions to preserve a contiguous batched layout as shown in Equation~\ref{eq:fc-layout}.
\begin{equation}
C_{\text{out}}^{(j)} \leftarrow \Rot\!\left(C_{\text{out}}^{(j)}, -j \cdot d_{\text{out}}\right).
\label{eq:fc-layout}
\end{equation}

Because encrypted batch neural networks generally operate on large batches of inputs, we introduce parameterization of this layer to reduce the number of rotation keys required. Instead of applying a single rotation by $-j \cdot d_{\text{out}}$ for every batch index, we decompose the rotation offset into two components as shown in Equation~\ref{eq:reduce-fc-index} and apply the corresponding rotations sequentially as as shown in Equation~\ref{eq:sub-fc-rotations}.

\begin{equation}
j \cdot d_{\text{out}} = \texttt{base\_idx} + \texttt{mod\_idx},
\label{eq:reduce-fc-index}
\end{equation}
\begin{equation}
\begin{aligned}
\Rot\!\left(C_{\text{out}}^{(j)}, -j \cdot d_{\text{out}}\right)
&=
\Rot\!\Bigl(
    \Rot\!\left(C_{\text{out}}^{(j)}, -\texttt{base\_idx}\right), \\
&\hspace{3.2em}
    -\texttt{mod\_idx}
\Bigr).
\end{aligned}
\label{eq:sub-fc-rotations}
\end{equation}

Each rotation is applied only when the corresponding index is non-zero, allowing the computation to use at most two rotation keys for a single index. 
This decomposition of rotation keys for alignment substantially reduces the number of rotation keys that must be generated and stored, while preserving the correctness of the batched output alignment.

Once all output blocks have been computed and aligned, the final output ciphertext is obtained by summing all intermediate ciphertexts and adding a repeated encoded bias vector as shown in Equation~\ref{eq:fc-final-layout}.
\begin{equation}
C_{\text{out}} = \sum_j C_{\text{out}}^{(j)} \;\oplus\; \mathbf{b}^{\text{enc}},
\label{eq:fc-output}
\end{equation}
where $\mathbf{b}^{\text{enc}}$ denotes the bias vector replicated across the batch layout.
The resulting ciphertext encodes the outputs as:
\begin{equation}
C_{\text{out}} = \Enc\left(
\mathbf{y}^{(0)} \mid \mathbf{y}^{(1)} \mid \cdots \mid \mathbf{y}^{(b-1)}
\right),
\label{eq:fc-final-layout}
\end{equation}
with each $\mathbf{y}^{(j)} \in \mathbb{R}^{d_{\text{out}}}$ occupying a contiguous block of slots.

This design evaluates the fully connected layer for all batch elements simultaneously while maintaining a compact ciphertext layout. 
By combining blockwise merging, rotation keys reuse, and batched inner-product evaluation, the proposed approach minimizes memory footprint and create data independent submodules for multi-threading.



%% file: sections/_multithreading.tex
\subsection{Multi-threading}
\label{subsec:multithreading}

In addition to algorithmic optimizations, we exploited thread-level parallelism by introducing a high degree of independence across the computation steps in both the convolutional and fully connected layers. This design enables efficient mapping to multi-core CPU architectures, thereby improving system resource utilization and overall throughput.

\subsubsection{Convolution Layer Multi-threading}
We decompose the convolution layer into two logically distinct submodules: \emph{input alignment} and \emph{convolution evaluation}. 
The input alignment submodule generates the set of spatially shifted ciphertexts as described in Section~\ref{sub:input_alignment} while the convolution evaluation submodule computes the convolution operation as decribed in Section~\ref{sub:convolution_evaluation}.
Since the same operations have to be computed across different ciphertexts they form great candidates for multi-threading. 

Let $T$ denote the number of available threads in a system. 
The set of concurrent evaluations in our design is $T$ disjoint subsets where each thread executes a unique operation.
In the input alignment phase, threads are independently assigned to the ciphertexts of the of the input vector. At the end,  disjoint subsets of the rotated subsets are aggregated for use in the convolution evaluation phase.
In the convolution evaluation phase, threads are also independently assigned to different output channels computations. Each thread uses an independent weights vector equivalent to a corresponding output channel. 
At the end of each submodule, the results from the different threads are aggregated into a vector corresponding to the correct output of that submodule.

When the required threads are more than the actual system threads, our design allows just a subset of concurrent executions. Once a thread's computation is completed, the thread is resigned until all independent outputs are computed. 
In this process, no synchronization is required except during the final aggregation step as we ensure that  no data dependencies exit between two concurrent threads. 
The only drawback of this approach is that it increases the memory requirements of the application as multiple threads execute concurrently with each thread requiring its independent computation resources.

\subsubsection{Fully Connect Layer Multi-threading}
\label{subsec:fc_multithreading}

For the fully connected layer, we exploit thread-level parallelism by decomposing the computation into two stages: \emph{per-output-channel evaluation} and \emph{batch-wise result placement}, as illustrated in Sections~\ref{sub:fc-evaluation} and~\ref{sub:output-reconstruction}. This decomposition exposes a high degree of parallelism while preserving a simple execution model. 

In the first stage, each output neuron is processed independently in a separate thread. Specifically, a thread multiplies the encrypted input ciphertext with the corresponding plaintext weight vector for a single output channel, followed by slot-wise accumulation to compute the partial encrypted output for all inputs in the batch. Since output channels are independent and do not share intermediate state, these computations is safely and efficiently executed in parallel across threads.

In the second stage, the intermediate results produced by different threads are reorganized to form the final batched output layout. Each thread is responsible for placing its partial results into the correct batch positions through a sequence of ciphertext rotations and additions. This stage is likewise parallelizable, as each thread operates on disjoint output regions and does not introduce data dependencies with other threads.

Synchronization between threads is only required during the final aggregation step where all per-channel contributions are combined into the output ciphertext vector. 
Throughout both stages, our design ensures the absence of write-write or read-write conflicts, enabling lock-free execution for the FC layer computation. As in the convolution layer, the primary cost of this multi-threaded approach is increased memory consumption, since each thread maintains its own intermediate ciphertexts as well as a vector to store output ciphertexts independently for all threads. However, this overhead is amortized by the substantial reduction in end-to-end latency achieved through this parallel execution.

%% file: sections/_otherlayers.tex
\subsection{Other Layers}

In addition to convolution and fully connected layers, convolution neural network inference requires nonlinear activation functions and pooling layers. In this work, we adopt the implementations provided by \fheon for both the ReLU activation function and GlobalPooling layers in our models. 
We extend these implementations to support vectorized inputs and parallelize the computation across all elements of the input vector. Since the operations on different vector elements are independent, they can be executed concurrently without introducing additional data dependencies. This parallelization improves throughput while preserving the original homomorphic circuit structure and security guarantees as designed in \fheon.

%% file: sections/Pipeline.tex
\section{Optimized Inference Pipeline}
\label{sec:pipeline}

Deep neural networks often employ downsampling via strided convolutions or pooling to reduce spatial dimensions of inputs. 
In SIMD ciphertext, they create empty slots which we can refill and increase our batch size. 
Figure~\ref{fig:basic_pipeline} illustrates the standard  ResNet-20 architecture used for HE-friendly inference. 
We will use this model's pipeline as an illustrative example in this work as all the concepts discuss generalized to most neural networks architectures. 

\begin{figure*}[t]
    \centering
    \includegraphics[width=\textwidth]{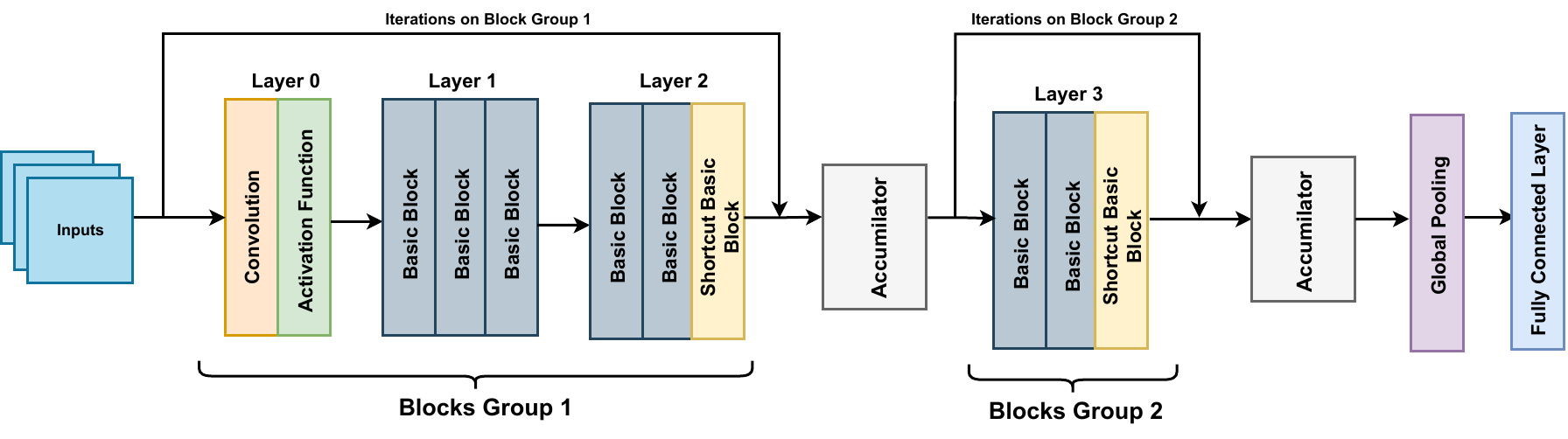}
    \captionsetup{justification=centering}
    \caption{Our Optimized ResNet-20 Pipeline with Accumulators. The Accumulators are used to rearranged and join the downsampled ciphertext output of the ResNet-20 Block Groups looping that precedes it}
    \label{fig:optimized_pipeline}
\end{figure*}


Let a ciphertext contain \(n\) slots, and an input feature map have spatial dimensions \(H \times W\) for a batch of size \(B\). The number of slots occupied per input ciphertext is shown in Equation~\ref{eq:slots-occupied}.
\begin{equation}
sts = H \cdot W \cdot B
\label{eq:slots-occupied}
\end{equation}

Our objective is to keep the value of $sts$ equal to the maximum number of slots available in the ciphertext throughout the HE-friendly model's evaluation pipeline.  
However, after downsampling with stride \(r_h \times r_w\), the spatial dimensions of the input reduces as shown in Equation~\ref{eq:reduce-dimensions}.
\begin{equation}
H' = \frac{H}{r_h}, \quad W' = \frac{W}{r_w},
\label{eq:reduce-dimensions}
\end{equation}

The total number of occupied slots in our SIMD ciphertext also reduces as shown in Equation~\ref{eq:slots-occupied-downsample}.

\begin{equation}
sts' = H' \cdot W' \cdot B = \frac{H \cdot W \cdot B}{r_h \cdot r_w}.
\label{eq:slots-occupied-downsample}
\end{equation}

The fraction of freed slots is shown in Equation~\ref{eq:freed-slots}.

\begin{equation}
f = 1 - \frac{sts'}{sts} = 1 - \frac{1}{r_h \cdot r_w},
\label{eq:freed-slots}
\end{equation}
Thus, the total number of free slots after downsampling  is shown in Equation~\ref{eq:number-slots},  where $n$ is the total number of available slots in the SIMD ciphertext. 
\begin{equation}
sts_\text{free} = n - sts'
\label{eq:number-slots}
\end{equation}

Rather than propagating partially filled ciphertexts through subsequent layers, our optimzed architectural design reclaims freed slots by packing multiple intermediary ciphertexts into a single ciphertext. 
Let \(g\) denote the number of independent iterations performed on the neural network's block groups before merging. 
We repeat the blocks up to a downsampling layer \(g\) times and then combine the intermediate ciphertexts to form a new ciphertext with an effective batch size increased by \(g\). 
We generally chose the values of $g$ such that $g$ multiplied by $sts'$ should be equal to $sts$.
The value of $g$ can be used to control the batch size of the encrypted inference. 
Merging is performed via masking and rotations as shown in Equation~\ref{eq:merge-ciphertexts}.

\begin{equation}
C_{\text{joined}} = \sum_{i=0}^{g-1} \mathrm{Rot}\left( C_i \odot M, -i \cdot sts' \right),
\label{eq:merge-ciphertexts}
\end{equation}

where \(C_i\) are intermediate ciphertexts, \(M\) is a plaintext cleaning mask, and \(\mathrm{Rot}(\cdot)\) denotes a slot rotation.

Figure~\ref{fig:optimized_pipeline} illustrates the optimized HE-friendly ResNet-20 pipeline used in this work. 
In this new architecture, we introduce an Accumulator to carryout the intermidiary ciphertext merging process.  
The accumulator reclaims and reuse freed slots from downsampling by merging multiple intermediate ciphertexts into a single, densely packed ciphertext before passing it for processing by subsequent layers. This enables a larger effective batch size, directly leads to a reduction in the amortized computation time.

Inference proceeds in a layer-wise pipelined manner, where inputs are processed in tiles, generating intermediate ciphertexts occupying disjoint slot regions. 
Ciphertexts are continuously merge by the accumulator once available as shown in Equation~\ref{eq:merge-ciphertexts}.  At the end of each block group loop, the network can proceed to  the next stage. 

To configure this architecture for different batch sizes, we first estimate the target effective batch size and divide it by the maximum number of input images that can be packed into a single encrypted input ciphertext. This ratio determines the required number of ciphertext merging operations after each downsampling block and serves as the iteration factor for the corresponding block groups accumulator. Using these parameters, we configure the inference pipeline accordingly giving us a very optimize way of setting different batch sizes while respecting the constraints in Equations~\ref{eq:slot-constraint} and \ref{eq:fc-slot-constraint}. 
The maximum number of ciphertext present in memory at any given time is equal to twice the total number of input channels for the next block group since ciphertexts are continuously merged at the end of every iteration.

\paragraph{Evaluation Keys Reuse.}
In the proposed pipeline, we adopt a key reuse strategy in which only the rotation and bootstrapping keys required for the current block group are loaded into memory. Within each block group, the spatial width of the encrypted inputs remains constant until a downsampling operation is applied. As a result, all blocks within the same group can reuse an identical set of rotation key up to the accumulator. Once the iteration is over, we offload these keys from memory and load a new set of keys for the subsequent block group. 
After the final block group, the keys are again offloaded and replaced with the keys required for the pooling and FC layers.
This staged key reuse strategy significantly reduces the overall memory footprint, as only the essential keys for the active computation stage are maintained in memory.

%% file: sections/Experiments.tex
\section{Experiments}

All Experiments were conducted on a system equipped with an Intel(R) Xeon(R) Gold 5418Y CPU, configured with 32 CPU cores, 256~GB of RAM, and running Rocky Linux~8.10. 
Each CPU core was configured to run a single hardware thread, resulting in a maximum of 32 concurrent threads.
We extended the \fheon framework using \openfhe v1.4.0 (released July~2025). These frameworks provide a comprehensive set of HE primitives offering a solid foundation for the development and evaluation of our system.
To the best of our knowledge, this will be the first open-source codebase for high-throughput encrypted neural network inference.

\subsection{HE Security Parameters}
\label{subsec:he-params}

All experiments are conducted using the CKKS scheme with  cryptographic context parameters selected to support efficient execution across HE-friendly CNN architectures. 
All setups use a multiplicative depth of $25$, a first modulus size of $50$ bits, a rescaling factor of $46$ bits, $4$ digits for key switching, and adopts the \texttt{flexibleauto} rescaling strategy. 
These parameters balance numerical stability, performance, and security, and remain fixed across all evaluation setups.

To evaluate different batch sizes using our proposed pipeline and implementations, we adopted two deployment settings that differ in polynomial degree and SIMD capacity of the ciphertexts.
These configurations are summarized in Table~\ref{tab:fhe_params}. 
Pipeline~1 uses a polynomial degree of  32,768, providing 16,384 SIMD slots for the pipeline ciphertexts. 
This number of slots allows up to 16 CIFAR-10 and CIFAR-100 images to be packed into a single input ciphertext for the HE-friendly neural network evaluation. 
Pipeline~2 doubles the polynomial degree to 65,536 and doubles the number of slots to  32,768 slots.
This configuration allows the input ciphertexts to hold up to 32 CIFAR-10 and CIFAR-100 images in the input ciphertexts. 

\begin{table}[http]
    \small
    \centering
    \caption{Pipelines parameters used for encrypted inference.}
    \label{tab:fhe_params}
    \begin{tabularx}{\linewidth}{|l|X|X|}
        \hline
        \textbf{Parameter} & \textbf{Pipeline 1} & \textbf{Pipeline 2} \\
        \hline
        Polynomial Degree    & 32,768 & 65,536 \\
        \hline
        Number of SIMD Slots & 16,384  & 32,768 \\
        \hline
    \end{tabularx}
\end{table}

The base batch sizes of $16$ for pipeline 1 and $32$ for pipeline 2 are expanded during inference by reclaiming freed slots at downsampling layers, as described in Section~\ref{sec:pipeline}. 
For the ResNet-20 architecture using CIFAR-10 dataset, which contain two spatial downsampling stages, the effective batch sizes grow multiplicatively as the pipeline progresses.
Based on this behavior, we evaluate three representative batch sizes per pipeline, shown in Table~\ref{tab:batchsizes}.

\begin{table}[http]
    \small
    \centering
    \caption{Evaluated batch sizes enabled by each pipeline.}
    \label{tab:batchsizes}
    \begin{tabularx}{\linewidth}{|X|X|}
        \hline
        \textbf{Pipeline} & \textbf{Batch Sizes} \\
        \hline
        Pipeline 1 & 16, 64, 256 \\
        \hline
        Pipeline 2 & 32, 128, 512 \\
        \hline
    \end{tabularx}
\end{table}

The base batch sizes of 16 and 32 correspond to a single straightforward pass in the HE-friendly neural networks, where the input ciphertexts are fully filled but no iterations on block groups are processed thus no merging of intermediate representations is required. These configurations minimize control overhead and favor inference scenarios where low overall runtime is required.
The intermediate batch sizes of 64 and 128 exploit the first downsampling stage. Freed slots created by spatial reduction are reclaimed by looping the pre-downsampling blocks four times and merging the resulting ciphertexts, yielding a \(4\times\) increase in effective batch size while preserving dense packing.
The largest batch sizes 256 and 512 fully saturate the optimized pipeline. 
In these HE-friendly neural networks, freed slots from all downsampling stages are recursively reclaimed, ensuring that ciphertexts remain densely packed throughout inference and maximizing amortization of expensive homomorphic operations thus inference time.
Sixteen iterations of the first block group are required while four iterations of the second block group are also required in the most sataurated pipeline.  

Beyond architectural efficiency, our configured batch sizes align with the most commonly used batch sizes seen in plaintext deep learning workloads. 
This natural alignment ensures that our encrypted pipelines remain both efficient and practically deployable without imposing artificial constraints on model execution.

\subsection{Models Evaluated}
\label{subsec:models}

We evaluate our work using different variants of the ResNet-20 and ResNet-34 architectures on the CIFAR-10 and CIFAR-100 datasets, respectively. While the ResNet-20 is a well-established benchmarks for encrypted inference, ResNet-34 also present complementary architectural characteristics that allow us to stress different aspects of this work.

ResNet-20 is a compact residual network with two spatial downsampling stages, making it a representative model for studying the effectiveness of slot reclamation and ciphertext merging in shallow-to-moderate depth networks. Its relatively small parameter count and widespread use in prior HE-based inference work enable direct comparison with existing approaches while showing the impact of our optimizations.
ResNet-34, in contrast, is substantially deeper and contains an additional downsampling stage, increasing both computational depth and ciphertext management complexity. Evaluating ResNet-34 allows us to assess the scalability of the proposed designs under deeper residual hierarchies and longer inference chains, where inefficient slot utilization would otherwise incur significant overhead.

For the ResNet-20 with CIFAR-10 dataset, we evaluated the two pipelines as described in Section~\ref{subsec:he-params}.
For ResNet-34 evaluated on CIFAR-100, the maximum supported batch size differs. Specifically, under pipeline~1, our largest batch size used is 128 images, whereas pipeline~2 supports up to 256 images. 
This limitation arises from the fact that the CIFAR-100 dataset containing 100 output classes.  
Under the slot constraint defined in Equation~\ref{eq:fc-slot-constraint}, it's FC layer can pack at most 163 images in pipeline~1 and 327 images in pipeline~2. To efficiently manage the iterative execution of the pipeline and to ensure that all ciphertexts processed in the initial block groups are fully utilized, we normalize the configuration to use eight iterations of the first block group and two iterations of the second block group.  
These settings also yield batch sizes that are commonly used in practical ML pipelines.

In total, we evaluated twelve variants of models (six each for ResNet-20 and ResNet-32), corresponding to different effective batch sizes.
Together, these evaluations demonstrate the adaptability of our work across different network depths, dataset complexities, and batch sizes, providing a comprehensive view of our approaches performance and scalability under realistic encrypted inference workloads.

All plaintext models were developed and trained in Python using PyTorch. 
During training, we used batch normalization after each convolutional layer. 
Batch normalization is used because it improves network training and keeps the data range within the network relatively small and stable \cite{ioffe2015batchnormalizationacceleratingdeep}. 
After training, the learned batch normalization weights were folded back into their respective convolutional kernel weights before exporting them \cite{EdouardYvinec-2022}. 
All weights were exported as CSV files and later imported into their HE-friendly equivalent model as supported in \fheon. 

%% file: sections/Results.tex
\section{Results}
\label{sec:results}

All results are reported in terms of end-to-end encrypted inference circuit time, total bootstrapping latency, memory overhead, and amortized inference latency per image.

\subsection{Accuracy}

All variants of a given HE-friendly neural network architecture achieved identical accuracy, yielding the same classification results. 
These results are also consistent with the results of the encrypted baseline reported for the single-input pipeline in \fheon. 
We ran the first 1024 images from the validation set of each dataset through each variant of our model. Table~\ref{tab:accuracy_comparison} summarizes the accuracies obtained from both architectures.

\begin{table}[http]
    \small
    \centering
    \caption{Accuracy of models in the Plaintext and FHE settings.}
    \label{tab:accuracy_comparison}
    \begin{tabularx}{\linewidth}{|X|X|X|X|}
        \hline
        \textbf{Architecture} & \textbf{Plaintext(\%)} & \textbf{Encrypted(\%)}  & \textbf{Images}\\
        \hline
        ResNet-20 & 92.8 & 92.2 & 1024 \\
        \hline
         ResNet-34   & 76.5 & 74.4 & 1024 \\
        \hline
    \end{tabularx}
\end{table}

\subsection{Performance Evaluation}

Tables~\ref{tab:high_throughput_networks} and~\ref{tab:high_throughput_networks_pipeline2} report the performance of Pipeline~1 and Pipeline~2, respectively, across different batch sizes for ResNet-20 and ResNet-34. We evaluate end-to-end latency (Lt), amortized latency per image (Amrt), and total batch processing time (Bts) to characterize both absolute execution cost and throughput efficiency under batched encrypted inference.

For both pipelines and model architectures, increasing the batch size results in higher total latency, reflecting the additional iterations over the batch-group required. 
However, the amortized latency consistently decreases as the batch size grows, indicating that our optimized pipeline effectively amortizes the the average runtime across multiple inputs. 

Pipeline~1 achieves lower absolute latency across all configurations, particularly for smaller batch sizes, due to it's faster homomorphic computations attributed to the smaller polynomial degree. 
Interestingly, using a ResNet-20 with a batch-size of $256$ results in an inference time of $8.35$, the smallest recorded in this work. 
In contrast, Pipeline~2 enables substantially larger batch sizes, which leads to improved amortized latency at scale. For ResNet-20, Pipeline~2 reduces the amortized latency from 29.25\, seconds at batch size 32 to 8.86\, seconds at batch size 512, approaching the best per-image performance observed in Pipeline~1 while supporting significantly higher throughput. 
A similar trend is observed for ResNet-34, where the amortized latency drops $96.25$  seconds to $25.1$ seconds and  from $102.96$ seconds to $28.14$ seconds as the batch sizes increases from $16$ to $128$ for pipeline 1 and from $32$ to $256$ in pipeline 2 respectively. 

Despite the higher total latency of Pipeline~2 at large batch sizes, the increase in total batch processing time remains well controlled and scales approximately linearly with the batch size. This indicates that the additional packing and merging operations introduced in Pipeline~2 do not incur superlinear overheads as the increase in polynomial degree is directly balanced by the number of images one can pack. 

\begin{table}[ht]
\centering
\caption{Results for Pipeline~1.}
\label{tab:high_throughput_networks}
\begin{tabular}{|l|c|c|c|c|}
\hline
\textbf{Model} & \textbf{Batch Size} & \textbf{Lt (s)} & \textbf{Amrt (s)} & \textbf{Bts (s)} \\
\hline
\multirow{3}{*}{ResNet-20} 
 & 16  & 442  & 27.625 & 1650 \\
\cline{2-5}
 & 64  & 691  & 10.796 & 2607 \\
\cline{2-5}
 & 256 & 2140 & 8.359  & 7989 \\
\hline
\multirow{3}{*}{ResNet-34} 
 & 16  & 1764 & 96.25 & 3628 \\
\cline{2-5}
 & 64  & 2348 & 36.678 & 4521 \\
\cline{2-5}
 & 128 & 3214 & 25.109 & 6343 \\
\hline
\end{tabular}
\end{table}

\begin{table}[ht]
\centering
\caption{Results for Pipeline~2.}
\label{tab:high_throughput_networks_pipeline2}
\begin{tabular}{|l|c|c|c|c|}
\hline
\textbf{Model} & \textbf{Batch Size} & \textbf{Lt (s)} & \textbf{Amrt (s)} & \textbf{Bts (s)} \\
\hline
\multirow{3}{*}{ResNet-20}
 & 32  & 936  & 29.25 & 2723 \\
\cline{2-5}
 & 128 & 1464 & 11.43 & 4483 \\
\cline{2-5}
 & 512 & 4540 & 8.86  & 14531 \\
\hline
\multirow{3}{*}{ResNet-34}
 & 32  & 3282 & 102.96 & 7229 \\
\cline{2-5}
 & 128 & 4982 & 38.92  & 8884 \\
\cline{2-5}
 & 256 & 7204 & 28.14  & 11908 \\
\hline
\end{tabular}
\end{table}

\subsection{Memory Usage}

Tables~\ref{tab:memory_pipeline1} and~\ref{tab:memory_pipeline2} summarize the memory requirements for Pipeline~1 and Pipeline~2 across different batch sizes for ResNet-20 and ResNet-34. Memory usage increases with batch size, as larger batches require more ciphertexts to be held in memory simultaneously within the pipelines. 

For Pipeline~1, ResNet-20 exhibits memory usage ranging from 43.42\,GB for a batch size of 16 to 48.88\,GB for a batch size of 256. 
ResNet-34 shows a similar trend but with higher memory requirements.
The small batch sizes of 16 requires 70.7\,GB  also increases linearly for larger batch sizes, reaching 98.74\,GB for batch size 128. 

\begin{table}[ht]
\centering
\caption{Memory usage for Pipeline~1.}
\label{tab:memory_pipeline1}
\begin{tabular}{|l|c|c|}
\hline
\textbf{Model} & \textbf{Batch Size} & \textbf{Memory Usage (GB)} \\
\hline
\multirow{3}{*}{ResNet-20} 
 & 16  & 43.42 \\
 \cline{2-3}
 & 64  &  47.36 \\
 \cline{2-3}
 & 256 & 79.65 \\
\hline
\multirow{3}{*}{ResNet-34} 
 & 16  & 68.69 \\
 \cline{2-3}
 & 64  & 73.34 \\
 \cline{2-3}
 & 128 & 94.89 \\
\hline
\end{tabular}
\end{table}


Pipeline~2 requires a larger polynomial degree to support larger batch sizes, resulting in higher memory consumption overall. 
ResNet-20 ranges from 87.01\,GB at batch size 32 up to 98.96\,GB at batch size 512. 
For ResNet-34, the memory grows more drastically from $157.23$ reaching 246.78\,GB for a batch size 256. 
The only models that requires more than 128~GB of RAM are the ResNet-34 pipeline 2 models. This increase is as a result of a combination of the deeper architecture and the number of classes in the CIFAR-100 dataset. 

\begin{table}[ht]
\centering
\caption{Memory usage for Pipeline~2.}
\label{tab:memory_pipeline2}
\begin{tabular}{|l|c|c|}
\hline
\textbf{Model} & \textbf{Batch Size} & \textbf{Memory Usage (GB)} \\
\hline
\multirow{3}{*}{ResNet-20} 
 & 32  & 87.01 \\
 \cline{2-3}
 & 128 & 95.81 \\
 \cline{2-3}
 & 512 & 98.96 \\
\hline
\multirow{3}{*}{ResNet-34} 
 & 32  & 156.68 \\
 \cline{2-3}
 & 128 & 196.77 \\
 \cline{2-3}
 & 256 & 246.78 \\
\hline
\end{tabular}
\end{table}

Overall, our analysis reveals a clear trade-off among batch size, model depth, throughput, and memory consumption. While Pipeline~2 consistently achieves higher throughput due to its support for larger effective batch sizes, its amortized latency for any given configuration is slightly higher than that of Pipeline~1. In addition, Pipeline~2 incurs a noticeable increase in memory usage at the base batch size, nearly doubling the memory footprint in some configurations.
This behavior is primarily driven by the increased polynomial degree of the underlying HE context. A higher polynomial degree enables packing a larger number of inputs per ciphertext, thereby improving throughput; however, it also increases ciphertext sizes, key sizes, and the computational cost of homomorphic operations. 
These trade-offs are inherent to all HE-based systems.

Consequently, the pipeline and batch size choices must be guided by the system constraints, including available memory, CPU core count, and desired latency-throughput balance. Our results provide concrete guidance for selecting configurations that best match practical deployment environments.

\subsection{Comparison with Related Works}
\label{subsec:related_comparison}

We compare our results against most recent works in high-throughput encrypted neural networks based on the ResNet-20 model and CIFAR-10 dataset. 
To contextualize our results, we then discuss our optimized Pipeline~2 against the state-of-the-art batch inference work by Cheon et al.~\cite{cheon2024batch} at a batch size of 512 CIFAR-10 images.
Cheon et al. achieved an amortized inference latency of 15.76\,s per image with a peak memory usage of 370.28\,GB. Their implementation was evaluated on a high-end system with 2$\times$ Intel Xeon Gold 6248 CPUs (40 cores in total) and 1\,TB of RAM.  
In contrast, our pipeline achieves an amortized latency of 8.86\,seconds per image, which is 1.78x reduction in inference time, while requiring only 98.96\,GB of memory, which is approximately 3.74x lower than their work. 
Furthermore, our we used an Intel(R) Xeon(R) Gold 5418Y CPU configured with 32 CPU cores and only 256 GB of RAM for our setup which is a much cheaper and smaller setup compared to any other system used in related works.
These results demonstrates that our proposed approach to high throughput encrypted neural networks  does not only reduce the per-image latency and memory footprint substantially but does so on a more cost-effective and realistic hardware.  
Table~\ref{tab:comparison_sota} compares our work with the state-of-the-art work of  Cheon et al.~\cite{cheon2024batch}

\begin{table}[http]
\small
\centering
\caption{Amortized inference latency and memory usage for ResNet-20 at batch size 512.}
\label{tab:comparison_sota}
\begin{tabularx}{\linewidth}{|l|X|X|}
\hline
\textbf{Method} & \textbf{Amortized (s)} & \textbf{Memory (GB)} \\
\hline
AutoFHE~\cite{ao2024autofhe} & 75 & --- \\
\hline
Cheon et al.~\cite{cheon2024batch} & 15.76 & 370.28 \\
\hline
Our Pipeline (Batch 512) & \textbf{8.86} & \textbf{98.96} \\
\hline
\end{tabularx}

\footnotesize\textit{Processor Summary:} \\
    AutoFHE: AWS Instances r5.24xlarge, 96 CPU cores, 768 GB RAM. \\
    Cheon et al.:  2$\times$ Intel Xeon Gold 6248 CPUs, 40 cores, and 1\,TB of RAM.  \\
    
    Our Work: Intel(R) Xeon(R) Gold 5418Y CPU, 32 cores, 256 GB of RAM\\
    
\end{table}

We also do not compare our results against single-input inference approaches such as \fheon~\cite{njungle2025fheon} or the work of Rovida et al.~\cite{rovida_cnn}. 
We consider both batched and single-input encrypted inference to be important but fundamentally different research directions, each with distinct strengths and application scenarios.

%% file: sections/Conclusion.tex
\section{Conclusion}
\label{sec:conclusion}

In this work, we presented highly efficient designs for high-throughput encrypted  neural network layers. We also presented an optimized pipeline for large-batch processing of HE-friendly neural networks. 
Furthermore, we carefully design our neural network layers to reduce data dependencies and apply multi-threading to further improve performances. 
Through extensive evaluation on ResNet-20 and ResNet-34 with CIFAR-10 and CIFAR-100 datasets, we demonstrated that our optimizations reduce amortized inference latency by $1.78\times$ and  and lower memory usage by $3.74\times$ compared to state-of-the-art methods. 
Our proposed pipeline also enables flexible batch sizes making PPML inference of different batch sizes practical within the encrypted domain.

Future work will explore extending these techniques to encrypted deep neural network training, which remains largely underexplored due to several significant challenges. 
Training deep neural networks under HE imposes significantly high computational and memory requirements when compared to inference.
This is because gradient-decent computation, backpropagation, and parameter updates must all be performed on encrypted data. 
Furthermore, training requires lots of iterations over the neural network layers for proper learning. 
Huge encrypted datasets are also required which takes up alot of storage and significantly increase latency(the training set of CIFAR-10 has 50,000 images). 
While our amortized inference pipeline achieves about 8 second in best case scenarios, training on encrypted data will also necessitate efficient design of all the other required operations as well as careful computational resource utilization.
We envision that batched homomorphic training could leverage similar slot-reclamation and multi-threading techniques to parallelize gradient computations across both batch elements and model parameters.